 \documentclass[authoryear,preprint,review,12pt]{elsarticle}
\usepackage{amsthm}
\usepackage{amsmath}
\usepackage{amssymb}
\usepackage{multirow}
\usepackage{natbib}
\usepackage{comment}
\usepackage[english]{babel}
\usepackage{setspace}
\usepackage{enumerate}
\usepackage{epstopdf}
\usepackage{color}
\usepackage{scalerel,stackengine}
\usepackage{mathtools}
\usepackage{url}
\makeatletter
\usepackage[margin=1 in]{geometry}
\usepackage{subfigure}
\usepackage{verbatim}
\usepackage{graphicx}
\usepackage{float}
\usepackage{bm}
\usepackage{esvect}
\theoremstyle{definition}
\newtheorem{definition}{Definition}
\newtheorem{theorem}{Theorem}
\numberwithin{equation}{section}
\newcounter{example}[section]
\newenvironment{example}[1][]{\refstepcounter{example}\par\medskip
   \noindent \textbf{Example~\theexample. #1} \rmfamily}{\medskip}

\usepackage{amsmath}

\begin{document}
	\doublespacing
	\begin{frontmatter}
\title{Modified Bivariate Weibull Distribution Allowing Instantaneous and Early Failures  }
%\author {Joseph Mathews\\ \texttt{jmathews6@my.apsu.edu}\thanks{Department of Mathematics and Statistics,
%		Austin Peay State University, Clarksville, TN.}
%	\and Sumen Sen\\ \texttt{sens@apsu.edu}\thanks{Department of Mathematics and Statistics, Austin Peay State University, Clarksville, TN.}
%	\and Ishapathik Das\\ \texttt{ishapathik@iittp.ac.in}\thanks{Department of Mathematics,
%		Indian Institute of Technology Tirupati, Tirupati,India.}}
	
%\begin{comment}	
\author[a]{Sumangal Bhattacharya}
\author[a]{Ishapathik Das\corref{c}}
\cortext[c]{Corresponding author}
\ead{ishapathik@iittp.ac.in}

\author[b]{Muralidharan Kunnummal}

\address[a]{Department of Mathematics, Indian Institute of Technology Tirupati, Tirupati, India.}
\address[b]{Department of Statistics, The M. S. University of Baroda, Vadodara, India.}
%\end{comment}
	
%\author{\small J.~Mathews$^a$, S.~Sen$^a$, I.~Das$^{c,}$\thanks{{Corresponding author. Email}:
%		ishapathik@iittp.ac.in, \vspace{6pt} {Phone}: 918772503357}\\
%	\small $^{a}$Department of Mathematics and Statistics, Austin Peay State University, Clarksville, TN.\\
%%	\small $^{b}$Department of Mathematical Sciences, University of Memphis, Memphis, USA\\
%	\small $^{c}$Department of Mathematics, Indian
%	Institute of Technology Tirupati, Tirupati, India}	

%\author{Joseph Mathews (jmathews6@my.apsu.edu) \hspace{5em} Sumen Sen (sens@apsu.edu)}

%\onecolumn

%\bibliographystyle{elsarticle-num}

%	\maketitle
	\begin{abstract}
In reliability and life data analysis, the Weibull distribution is widely used to accommodate more data characteristics by changing the values of the parameters. We frequently observe many zeros or close to zero data points in reliability and life testing experiments. We call this phenomenon a nearly instantaneous failure. Many researchers modified the commonly used univariate parametric models such as exponential, gamma, Weibull, and log-normal distributions to appropriately fit such data having instantaneous failure observations. 
 %%To encounter this situation, researchers proposed modified Weibull distribution by a mixture of two parameters Weibull distribution and a singular distribution at zero for univariate data.%% 
 Researchers also find bivariate correlated life testing data having many observations near a point $(x_0,y_0)$ while the remaining observations follow some continuous distribution. This situation defines as responses having early failures for such bivariate responses. In particular, if the point is the origin [i.e., $(x_0,y_0)=(0,0)$], then we call the situation a nearly instantaneous failure for the responses. Here, we propose a modified bivariate Weibull distribution that allows early failure by combining bivariate uniform distribution and bivariate Weibull distribution. The bivariate Weibull distribution is constructed using a 2-dimensional copula, assuming the marginal distributions as two parametric Weibull distributions. We derive some properties of that modified bivariate Weibull distribution, mainly the joint probability density function, the survival (reliability) function, and the hazard (failure rate) function. The model's unknown parameters are estimated using the Maximum Likelihood Estimation (MLE) technique combined with a machine learning clustering algorithm. Numerical examples are provided using simulated data to illustrate and test the performance of the proposed methodologies. The method is also applied to real data and compared with existing approaches to model such data in the literature. 
	\end{abstract}


\end{frontmatter}
	\section{Introduction}\label{intro}
In reliability analysis and life data analysis, Weibull distribution is used mostly because of its versatility.
In literature, extension and generalization of the Weibull distribution are given in
%to deal such univariate life  data% 
 \cite{pham2007Onrecent},  \cite{murthy2004weibull},  \cite{elgohary2015bivariate}. 
In life testing experiments, mainly in electronic parts and clinical trials, many items fail instantaneously or fail early due to faulty construction or inferior quality or due to no response of the treatments. We record the observed lifetime is close to zero. To model life data containing such nearly instantaneous failure, traditional parametric distributions, for example, Weibull distribution, Pareto distribution, exponential distribution, gamma distribution, log-normal distribution, etc., are not accurate in practice as such failures in experiment usually discard the assumption of a uni-modal distribution. These situations can be modeled by modifying the traditionally used parametric distributions.
\cite{muralidharan2006analysis}, modified the two parameters Weibull distribution by mixing two-parameter Weibull distribution and a singular distribution at zero.
%%But there is some problem to defining survival (failure rate) function due to this singular distribution.%%
\cite{lai2007weibull} modified the two parameters Weibull distribution by considering the mixture of the uniform distribution and two parameters Weibull distribution.
\cite{muralidharan2011modified}, model this instantaneous or early failures data by considering the mixture of the uniform distribution and the Pareto distribution. To estimate the model parameters, maximum likelihood estimation is used mainly.  \cite{muralidharan2011bayesian}, estimated the model parameters using Bayesian approach. \cite{cheng2016robust} proposed M-estimation method and derived the closed form of the shape parameter estimation. Using simulation study they showed that, M-estimation performs  better than the MLE for samples with early failure. 
To model data with inliers (instantaneous or early failure), non-standard mixture of distributions, with degeneracy occurring at zero and one, and a probability distribution for positive observation is used  mainly. \cite{muralidharan2018new} estimates the parameters of such model on the basis of type-II censored sample from a Weibull distribution with discrete mass at zero and one. 
To model instantaneous failure data from clinical trials and longitudinal,   \cite{withana2020extended} proposed two methods to estimate the parameters and used expectation-maximization(EM) algorithm. 
\cite{muralidharan2021some} provided a method for testing hypothesis to find statistical significance of the single and multiple inliers %(instantaneous or early failures)
in samples.  

Many data in real life are bivariate in nature and correlated, which contains huge observation near a particular point $(x_0,y_0)$ and other responses follow some continuous distribution; we define this as early failure data. If that particular point is origin, i.e. $(x_0,y_0)=(0,0)$, we call it as nearly instantaneous failure data. Bivariate Weibull distribution using FGM copula and there some properties are proposed in literature \cite{almetwally2020bivariate},  \cite{peres2018bivariate}. However, we did not find any modeling technique to fit such correlated bivariate data with instantaneous or nearly instantaneous failure observations adequately in the literature. Here, we propose a modified bivariate Weibull distribution allowing early failure by considering the mixture of the bivariate uniform distribution and bivariate  Weibull distribution. The bivariate Weibull distribution is obtained using a copula and two parameters Weibull distribution as marginal distribution. The forms of the survival function and hazard function are obtained by using a generalization of the Farlie-Gumbel-Morgenstern (GFGM) copula given by \cite{lai2000new}.
Furthermore, we consider the model to deal with the presence of nearly instantaneous failure $[(x_0,y_0)=(0,0)]$ in the responses and obtained the form of the hazard function and survival function using the GFGM copula. 
To illustrate the proposed model, in the simulation study, we consider the Gaussian copula [\cite{nelsen2007introduction}] to construct the modified bivariate Weibull distribution. To estimate the parameters, we use the method of maximum likelihood estimation (MLE) and density-based spatial clustering of applications with noise (DBSCAN) clustering given by \cite{ester1996density}.
%In the numerical study, we estimate real-life bivariate data.
%Here, in section 2, we review the generalized bi-variate Dirac delta distribution, two parameters Weibull distribution, the copula function with some example and the Sklar's theorem; in section 3, we find the bi-variate Weibull distribution using copula having margins univariate two parameters Weibull distribution and form of the distribution using  FGM copula;
%in section 4, we derive the form of the modified Weibull distribution by mixing the generalized bivariate Dirac delta distribution and the bivariate Weibull distribution and find the form of Hazard function and survival function using generalized FGM copula; in section 5, we discuss the case of nearly instantaneous failure and find the form of the survival function ; 
%in section 6, we discuss the MLE as a  method of estimation; in section 7, we have given one algorithm to simulate data; in  subsection 7.1, we simulate data and apply the proposed  model and estimate the parameters by MLE estimation method and checked the goodness of fit; in subsection 7.2, we use the proposed method to the bivariate real-life data; 
%in section 8, we have given the conclusion.

\par The remainder of the article is organized as follows. We provide a brief review of copula functions and Sklar's theorem for constructing multivariate distribution functions using the known marginals in Section \ref{copulasklar}. In Section \ref{bwd}, we construct a bivariate Weibull distribution function using a copula having two parameters Weibull distributions as marginals. We propose a modified Weibull distribution for early and instantaneous failure bivariate observations along with their hazard functions and the survival functions in Sections \ref{mbwd} and \ref{nearlyinstfailure}. Estimation of unknown parameters using the MLE combined with DBSCAN clustering process is described in Section \ref{estimation}. In Section \ref{algorithm}, we provide an algorithm to simulate data from modified bivariate Weibull distribution. Numerical examples with simulated data are given in Section \ref{simulation}, and the proposed method is applied to real data in \ref{real_data}. The concluding remarks are given in Section \ref{conclusion}.

	\section{Copulas and Sklar's Theorem}\label{copulasklar}
	In this section, copula function and how it can be used to form multivariate joint probability mass function from its marginals are discussed. For more information about copula we refer \cite{nelsen2007introduction}, \cite{joe2014dependence} and \cite{song2007correlated}.
	
	%In this section, we describe how the copula function may be used to form a multivariate probability mass function. We refer to  \cite{joe2014dependence} and \cite{song2007correlated} for more information on copulas than presented here. A copula is a multivariate probability distribution with uniform marginal distributions on the interval $[0,1]$.
	
	\begin{definition}\label{def1}
		A d-dimensional copula C is a function from $[0,1]^{d}$ to $[0,1]$ satisfying the following conditions:
		\begin{enumerate}
			\item $C(1, \ldots, 1,y_{j}, 1,  \ldots, 1)  = y_{j}, \forall \ j=1,2,\ldots,d$ where  $(1, \ldots, 1, y_{j}, 1, \ldots, 1) \in [0,1]^d$,
			\item $C(y_{1},y_{2}, \ldots, y_{d}) = 0$ if at least one $y_{j} = 0$ for $j = 1, 2, \ldots, d$, 
			\item For any $\textbf{y}_1=(y^{(1)}_{1}, y^{(1)}_{2}, \cdots , y^{(1)}_{d})$, $\textbf{y}_2=(y^{(2)}_1, y^{(2)}_2, \cdots , y^{(2)}_d)$ belong to $[0,1]^d$ with $y^{(1)}_j \leq y^{(2)}_j$, for all $j = 1, 2, \ldots, d$:
			\begin{equation}\label{eq2.1}
				\begin{aligned}
					\sum_{i_{1} = 1}^{2}\sum_{i_{2} = 1}^{2} \ldots \sum_{i_{d} = 1}^{2} (-1)^{i_{1} + i_{2} + \ldots + i_{d}}C(y^{(i_1)}_{1},y^{(i_2)}_{2}, \ldots, y^{(i_d)}_{d} )\geq 0 .
				\end{aligned}
			\end{equation}
		\end{enumerate}
	\end{definition}
	In copula theory, Sklar's theorem [\cite{sklar1959fonctions}] is a fundamental result and we can use it to form a joint distribution from its marginal distributions. 
	\begin{theorem}\label{thm1}
		Let  $F_{1}, F_{2}, \ldots, F_{d}$ are the marginal distribution functions for the random variables $Y_{1}, Y_{2}, \ldots, Y_{d}$ and their joint cumulative distribution function be F, then the following are true:
		\begin{enumerate}
			\item $\exists$ a d-dimensional copula C s.t., $\forall$ $(y_{1}, y_{2}, \ldots, y_{d}) \in \mathbb{R}^d$,
			\begin{equation}\label{eq2.2}
				\begin{aligned}
					F(y_{1}, y_{2}, \ldots, y_{d}) = C(F_{1}(y_{1}), F_{2}(y_{2}), \ldots, F_{d}(y_{d})) ,
				\end{aligned}
			\end{equation}
			\item If $Y_{1}, Y_{2}, \ldots, Y_{d}$ are continuous then the copula C is unique. Otherwise, C can be uniquely determined on a d-dimensional rectangle $Range(F_{1}) \times Range(F_{2}) \times \ldots \times Range(F_{d}).$ 
		\end{enumerate}
	\end{theorem}
 A copula is a multivariate probability distribution with uniform marginal distributions on the interval $[0,1]$. 
Copulas are popular because of their ability to model dependence among $Y_{1}, Y_{2}\cdots, Y_{d}$. Here is some examples of copula

\begin{example}
The multivariate Gaussian copula is given by the function:
		\begin{equation}\label{eq2.3}
		\begin{aligned}
		C(y_{1}, y_{2}, \ldots, y_{d}|R) = \bm{\Phi}_{R}(\Phi^{-1}(y_{1}), \Phi^{-1}(y_{2}), \ldots, \Phi^{-1}(y_{d})), 
		\end{aligned}
		\end{equation}
		where $(y_{1}, y_{2}, \ldots, y_{d})\in [0,1]^d$,  $\Phi^{-1}$ is the inverse of cumulative distribution function of the standard Gaussian distribution and $\bm{\Phi}_{R}$ is the joint CDF of a standard multivariate Gaussian distribution with correlation matrix $R$.
		
		The Gaussian copula density (\cite{arbenz2013bayesian}) is given by
		\begin{equation}\label{eq2.4}
			\begin{aligned}
				c(y_{1}, y_{2}, \ldots, y_{d} \mid  R) = \frac{1}{\sqrt{\mid R\mid }} exp \left\{-\frac{1}{2}\textbf{U}^{T} \times \left(R^{-1} - I_{d}\right) \times \textbf{U} \right\}  ,
			\end{aligned}
		\end{equation}
		where $\textbf{U} = \left[\Phi^{-1}(y_{1}), \Phi^{-1}(y_{2}), \ldots, \Phi^{-1}(y_{d})\right]^{T}$, $(y_{1}, y_{2}, \ldots, y_{d})\in [0,1]^d$, $\Phi$ is the univariate standard normal distribution and $R$ is the correlation matrix of the standard multivariate Gaussian distribution and $I_d$ be the $d\times d$ identity matrix.
\end{example}

\begin{example}
A generalization of the bivariate  Farlie-Gumbel-Morgenstern (GFGM) copula \cite{lai2000new} is given by 
\begin{equation}\label{fgmcopula}
    C(x,y)=xy+\rho x^by^b(1-x)^a(1-y)^a
\end{equation}
where $(x,y)\in [0,1]^2$, $a\geq1,\ b\geq1$ and $-1\leq \rho \leq1$. 
\end{example}

Therefore, the density function of the GFGM copula is given by
\begin{equation}\label{fgmcopula}
\begin{split}
    c(x,y)=&1+b^2\rho x^{b-1}y^{b-1}(1-x)^a(1-y)^a-ab\rho x^{b-1}y^b(1-x)^a(1-y)^{a-1}-\\
    &ab\rho x^by^{b-1}(1-x)^{a-1}(1-y)^a
    +a^2\rho x^b y^b(1-x)^{a-1}(1-y)^{a-1}
\end{split}
\end{equation}
where $(x,y)\in [0,1]^2$, $a\geq1,\ b\geq1$ and $-1\leq \rho \leq1$.

	\section{Bivariate  Weibull Distribution}\label{bwd}
	Let us recall the definition of two parameters Wibull distribution. 
	We denote $W\sim Wb(\alpha,\beta)$, if a random variable $W$ follows Weibull distribtuion with parameters $(\alpha,\beta)$ with the CDF given by
	\begin{equation}\label{univariatecdf}
    F_W(x)=1-\exp\left[-\left(\dfrac{x}{\beta}\right)^{\alpha} \right], \quad \quad x >0,\alpha>0,\beta>0
\end{equation}
	and the PDF given by
\begin{equation}\label{univariatepdf}
    f_W(x)=\dfrac{\alpha}{\beta}\left(\dfrac{x}{\beta}\right)^{\alpha-1}\exp\left[-\left(\dfrac{x}{\beta}\right)^{\alpha} \right], \quad \quad x >0,\ \alpha>0,\ \beta>0
\end{equation}

Let $X\sim Wb(\alpha_1,\beta_1)$ and $Y\sim Wb(\alpha_2,\beta_2)$ follow two parameter Weibull distribution with parameters  $( \alpha_1, \beta_1)$ and $( \alpha_2, \beta_2)$ respectively with $ \alpha_1, \beta_1 , \alpha_2, \beta_2 >0$. 
Therefore the joint CDF of the bivariate Weibull distribution using a  2-dimensional copula $C$ and Sklar's theorem is given by 
\begin{equation}\label{bivariateweibullcdfuc}
    F_{XY}(x,y) = C(F_X(x),F_Y(y))
\end{equation}
where $x>0$, $y>0$ and $F_X, F_Y$ are the CDF  of the random variables X and Y respectively. %which are given by 
%\begin{equation}\label{univariatecdf}
%    F_X(x)=1-\exp\left[-\left(\dfrac{x}{\beta_1}\right)^{\alpha_1} \right], \quad \quad x >0,\alpha_1>0,\beta_1>0
%\end{equation}
%and \textcolor{red}{$F_Y$ also %have the similar expression}.

The joint probability density function (PDF) of the bivariate  Weibull distribution is given by
\begin{equation}\label{bivariateweibullpdfuc}
    f_{XY}(x,y) = f_X(x)f_Y(y)\dfrac{\partial^2 C(F_X(x),F_Y(y))}{\partial F_X \partial F_Y}
\end{equation}
where $x>0$, $y>0$ and $f_X(x),\  f_Y(y)$ are the PDF of the random variables X and Y respectively.
%which are given  by
%\begin{equation}\label{univariatepdf}
%    f_X(x)=\dfrac{\alpha_1}{\beta_1}\left(\dfrac{x}{\beta_1}\right)^{\alpha_1-1}\exp\left[-\left(\dfrac{x}{\beta_1}\right)^{\alpha_1} \right], \quad \quad x >0,\ \alpha_1>0,\ \beta_1>0
%\end{equation}
%and \textcolor{red}{$f_Y(y)$ have the similar expression} and $C$ is a bivariate copula.\\
%\textcolor{red}{ Is different section needed or we can write it in different way ?}
%\subsection{Joint CDF and PDF using GFGM copula}\label{pdfcdfgfgm}\\

Now, we consider the GFGM copula \eqref{fgmcopula} to find the form of the joint CDF and joint PDF of the bivariate Weibull distribution. For simplicity of the expressions, we define notation for some commonly used term  as\\
\begin{equation}
    \begin{split}
        A=&\left(\dfrac{x}{\beta_1} \right)^{\alpha_1}\\
        B=&\left(\dfrac{y}{\beta_2} \right)^{\alpha_2}\\
        C=&A+B\\
        D=&\exp\left(-(a-1)C  \right)\left((a+b)\exp\left(-A \right)-a \right) \left((a+b)\exp\left(-B \right)-a \right).
    \end{split}
    \label{ABCD}
\end{equation}

%where $x>0$, $y>0$, $\alpha_1>0$, $\beta_1>0$, $\alpha_2>0$, $\beta_2>0$, $a\geq 1$, $b\geq 1$.

Therefore, the joint CDF of the bivariate Weibull distribution using the GFGM copula \eqref{fgmcopula}  is given by 
\begin{equation}\label{bivariateweibullcdfg}
    F_{XY}(x,y) = F_X(x)F_Y(y)+\rho  F_X(x)^bF_Y(y)^b (1-F_X(x))^a(1-F_Y(y))^a
\end{equation}
where $x>0$, $y>0$, $\alpha_1>0$, $\beta_1>0$, $\alpha_2>0$, $\beta_2>0$, $a\geq 1$, $b\geq 1$, $-1\leq \rho \leq 1$ and $F_X(x)$, $F_Y(y)$ are the CDF of $X$ and $Y$ respectively.\\
Therefore, the joint PDF $f_{XY}$ of the bivariate Weibull distribution is given by
%\begin{equation}\label{bivariateweibullpdfgeneral}
%\begin{split}
%    f_{X,Y}(x,y)= &\dfrac{\alpha_1 \alpha_2}{\beta_1 \beta_2}\left(\dfrac{x}{\beta_1} \right)^{\alpha_1-1}\left(\dfrac{y}{\beta_2} \right)^{\alpha_2-1}\exp\left[-\left(\dfrac{x}{\beta_1} \right)^{\alpha_1}-\left(\dfrac{y}{\beta_2} \right)^{\alpha_2} \right]\times\\
%    &\left[1+\rho \left(1-\exp\left(-\left(\dfrac{x}{\beta_1} \right)^{\alpha_1}\right)\right)^{(b-1)}\left(1-\exp\left(-\left(\dfrac{x}{\beta_2} \right)^{\alpha_2}\right)\right)^{(b-1)} \right. \times\\
%    &\exp\left(-(a-1)\left(\left(\dfrac{x}{\beta_1} \right)^{\alpha_1}+\left(\dfrac{y}{\beta_2} \right)^{\alpha_2} \right) \right)\left((a+b)\exp\left(-\left(\dfrac{x}{\beta_1}\right)^{\alpha_1} \right)-a \right)\times\\
%    &\left. \left((a+b)\exp\left(-\left(\dfrac{y}{\beta_2}\right)^{\alpha_2} \right)-a \right)\right]
%     \end{split}
%\end{equation}
%\begin{equation}\label{bivariateweibullpdfgeneral}
%\begin{split}
%   f_{X,Y}(x,y)= &\dfrac{\alpha_1 \alpha_2}{\beta_1 \beta_2}A^{1-1/\alpha_1}B^{1-1/\alpha_2}\exp(-C)
%   [1+\rho \left(1-\exp\left(-A\right)\right)^{(b-1)} \times\\
%   &\left(1-\exp\left(-B\right)\right)^{(b-1)}\exp\left(-(a-1)C  \right)\left((a+b)\exp\left(-A \right)-a \right)\times\\
%     &\left. \left((a+b)\exp\left(-B \right)-a \right)\right]
%    \end{split}
%\end{equation}
\begin{equation}\label{bivariateweibullpdfgeneral}
\begin{split}
    f_{XY}(x,y)= &\dfrac{\alpha_1 \alpha_2}{\beta_1 \beta_2}A^{1-1/\alpha_1}B^{1-1/\alpha_2}\exp(-C)
    [1+\rho \left(1-\exp\left(-A\right)\right)^{(b-1)}
    \left(1-\exp\left(-B\right)\right)^{(b-1)}D]
     \end{split}
\end{equation}
where $x>0$, $y>0$, $\alpha_1>0$, $\beta_1>0$, $\alpha_2>0$, $\beta_2>0$, $a\geq 1$, $b\geq 1$, $-1\leq \rho \leq 1$.
%and $A$, $B$, $C$, and $D$ are given in \eqref{ABCD} (\textcolor{red}{ should I have to write like this in each equation where A,B,C and D used ?}).

By taking restrictions on $a$ and $b$, we can get a simplified form of the bivariate Weibull distribution functions.
\begin{itemize}
\item If we choose $a=1,b=1$, the form of the joint CDF is  
\begin{equation}\label{bivariateweibullcdfp}
   \begin{split}
    F_{XY}(x,y) = &   F_X(x)F_Y(y)[1+\rho(1-F_X(x))(1-F_Y(y))]\\
     =&\left[1-\exp\left(-A \right) \right] \left[1-\exp\left(-B \right) \right]
   \left[1+\rho\exp\left[-\left(C\right) \right] \right]
   \end{split}
\end{equation}
where $x>0$, $y>0$, $\alpha_1>0$, $\beta_1>0$, $\alpha_2>0$, $\beta_2>0$, $0\leq \rho \leq 1$.\\
%and A,B,C are defined in equation \eqref{bivariateweibullpdfgeneral}.\\
Therefore the corresponding joint PDF of the bivariate Weibull distribution is
\begin{equation}\label{bivariateweibullpdfpp}
\begin{split}
    f_{XY}(x,y)= &\dfrac{\alpha_1 \alpha_2}{\beta_1 \beta_2}A^{1-1/\alpha_1}B^{1-1/\alpha_2}\exp(-C)\left[1+\rho \left(2\exp\left(-A \right)-1 \right) \left(2\exp\left(-B \right)-1 \right)\right]
     \end{split}
\end{equation}
where $x>0$, $y>0$, $\alpha_1>0$, $\beta_1>0$, $\alpha_2>0$, $\beta_2>0$, $-1\leq \rho \leq 1$.
\end{itemize}

	%\newpage
	\section{Modified Bivariate Weibull Distribution}\label{mbwd}
	Here we find the distribution function of the modified Weibull distribution and the form of the survival function and hazard function.
	\subsection{PDF and CDF of modified bivariate Weibull distribution }
We consider the modified bivariate model allowing the early  failure,  as a mixture of the bivariate uniform distribution and the bivariate Weibull distribution. Thus the joint PDF of the  modified bivariate Weibull (MBW) distribution is of the form
\begin{equation}\label{modifiedpdf}
    \begin{split}
        f(x,y) & =pf_1(x,y)+qf_2(x,y)\\
    \end{split}
\end{equation}
where $p+q=1$, $0<p<1$, $f_1(x,y)$ be the bivariate uniform distribution on the rectangle $[t_0, t_0+d]\times [t_0, t_0+d]$, given by   
\begin{equation}\label{bivariatedirac}
 f_1(x,y)= \left\{
        \begin{array}{ll}
            \dfrac{1}{d^2}, & \quad  x_{0} \leq x \leq x_{0}+d,\  y_{0} \leq y \leq y_{0}+d \\
            0, & \quad \text{Otherwise}
        \end{array}
    \right.
\end{equation}
for sufficiently small $d(>0)$, %$(x,y)\in \mathbb{R}\times\mathbb{R}$ %and the point $(x_0,y_0)\in \mathbb{R}^+\times \mathbb{R}^+$
%\textcolor{red}{($x_0 (\geq 0)$, $y_0 (\geq 0)$, should I give like that?)} 
and $f_2(x,y)=f_{XY}(x,y)$ is the bivariate Weibull distribution, given in \eqref{bivariateweibullpdfuc}.
\\
Therefore, the joint CDF of the MBW distribution is,
\begin{equation}\label{modifiedcdf}
    F(x,y)=pF_1(x,y)+qF_2(x,y)
\end{equation}
where $F_1$ and $F_2$ are joint CDF of the  bivariate uniform distribution and the bivariate Weibull distribution respectively.
\subsection{Survival function and Hazard  function }\label{survivalhazardnew}
The survival function of the MBW distribution is given by 
\begin{equation}\label{surviaval}
   \begin{split}
    R(x,y) &= F((X,Y)>(x,y))\\
    &=pF_1((X,Y)>(x,y))+qF_2((X,Y)>(x,y))\\
    &= pR_1(x,y)+qR_2(x,y)
   \end{split}
\end{equation}
where $R_1$ and $R_2$ are the survival functions of the  bivariate uniform distribution and the bivariate Weibull distribution respectively.\\
The hazard function of the MBW distribution is  given by
\begin{equation}\label{hazardfun}
    h(x,y)=\dfrac{\dfrac{\partial^2R(x,y)}{\partial x \partial y}}{R(x,y)}=\dfrac{f(x,y)}{R(x,y)}.
\end{equation}
Now, the survival functions $R_1$, $R_2$ and the hazard functions $h_1$, $h_2$ of the respective component distributions are given respectively
\begin{equation}\label{survivaldirac}
 R_1(x,y) =\left\{
        \begin{array}{ll}
            1, & \quad  x\leq x_{0},\ y \leq y_{0} \\
            \dfrac{x_0+d-x}{d}, & \quad x_0\leq x \leq x_0+d,\  y \leq y_0\\
            \dfrac{y_0+d-y}{d}, & \quad x\leq x_0 ,\  y_0 \leq y \leq y_0+d\\
             \dfrac{(x_0+d-x)(y_0+d-y)}{d^2}, & \quad  x_{0} \leq x \leq x_{0}+d,\ y_{0} \leq y \leq y_{0}+d\\
             0, & \quad  x_0+d \leq x \ \text{or}\  y_0+d \leq y
        \end{array}
    \right.
\end{equation}

\begin{equation}\label{hazarddirac}
 h_1(x,y) =\left\{
        \begin{array}{ll}
            \dfrac{1}{(x_0+d-x)(y_0+d-y)}, & \quad   x_{0} \leq x < x_{0}+d,\ y_{0} \leq y < y_{0}+d\\
            \infty, & \quad  x_0+d \leq x \ \text{or}\  y_0+d \leq y\\
             0, & \quad  \text{otherwise}
        \end{array}
    \right.
\end{equation}

\begin{equation}\label{survivalweiullgc}
      \begin{split}
        R_2(x,y)=& 1-F_X(x)-F_Y(y)+F_{XY}(x,y)\\
                =& 1-F_X(x)-F_Y(y)+C(F_X(x),F_Y(y))
      \end{split}
\end{equation}
Where $F_X$, $F_Y$, and $F_{XY}$ are the CDF of the random variables X, Y, and the bivariate Weibull distribution, respectively, and $C$ is a bivariate copula.

\begin{equation}
    h_2(x,y)=\dfrac{f_{X,Y}(x,y)}{R_2(x,y)}
\end{equation}
where the joint PDF of the bivariate Weibull distribution $f_{XY}(x,y)$ is given in \eqref{bivariateweibullpdfuc} and the survival function $R_2(x,y)$ is given by \eqref{survivalweiullgc}.

In particular, if we consider copula as GFGM copula, then the survival function $R_2(x,y)$ and hazard function $h_2(x,y)$ of the bivariate Weibull distribution is given by
%\begin{equation}\label{survivalweiullfgm}
%  \begin{split}
%        R_2(x,y)=&1-F_X(x)-F_Y(y)+F_{X,Y}(x,y)\\
%          =&1-F_X(x)-F_Y(y)+F_X(x)F_Y(y)+\rho F_X^b(x)F_Y^b(y)\times\\
%          & (1-F_X(x))^a(1-F_Y(y))^a\\
%          = &\exp\left[-\left(\left(\dfrac{x}{\beta_1} \right)^{\alpha_1} + \left(\dfrac{y}{\beta_2} \right)^{\alpha_2} \right) \right]\times\\
%          & \left(1+\rho\exp\left[-(a-1)\left(\left(\dfrac{x}{\beta_1} \right)^{\alpha_1}+ \left(\dfrac{y}{\beta_2} \right)^{\alpha_2}\right)\right]\right.\times\\
%          &\left.\left(1-\exp\left[-\left(\dfrac{x}{\beta_1}\right)^{\alpha_1} \right] \right)^b \left(1-\exp\left[-\left(\dfrac{y}{\beta_2}\right)^{\alpha_2} \right]\right)^b\right)
%  \end{split}
%\end{equation}
\begin{equation}\label{survivalweiullfgm}
  \begin{split}
        R_2(x,y)=&\exp\left(-C \right) \left[1+\rho\exp\left(-(a-1)C\right)\left(1-\exp\left(-A \right) \right)^b \left(1-\exp\left(-B \right)\right)^b\right]
  \end{split}
\end{equation}
where $x>0$, $y>0$, $\alpha_1>0$, $\beta_1>0$, $\alpha_2>0$, $\beta_2>0$, $a\geq 1$, $b\geq 1$, $-1\leq \rho \leq 1$.

and
% \begin{equation}\label{hazardweibullfgm}
% \begin{split}
% h_2(x,y) = &\dfrac{\begin{split}
%     \dfrac{\alpha_1 \alpha_2}{\beta_1 \beta_2}\left(\dfrac{x}{\beta_1} \right)^{\alpha_1-1}\left(\dfrac{y}{\beta_2} \right)^{\alpha_2-1}\exp\left[-\left(\dfrac{x}{\beta_1} \right)^{\alpha_1}-\left(\dfrac{y}{\beta_2} \right)^{\alpha_2} \right]\times\\ 
%     \left(1+\rho \left(1-\exp\left(-\left(\dfrac{x}{\beta_1} \right)^{\alpha_1}\right)\right)^{(b-1)}\left(1-\exp\left(-\left(\dfrac{x}{\beta_2} \right)^{\alpha_2}\right)\right)^{(b-1)}\right.\times \\
%     \exp\left[-(a-1)\left(\left(\dfrac{x}{\beta_1} \right)^{\alpha_1}+\left(\dfrac{y}{\beta_2} \right)^{\alpha_2} \right) \right]\times\\ \left.\left((a+b)\exp\left[-\left(\dfrac{x}{\beta_1}\right)^{\alpha_1} \right]-a \right) \left((a+b)\exp\left[-\left(\dfrac{y}{\beta_2}\right)^{\alpha_2} \right]-a \right)\right)\end{split}}{\begin{split}
%        \exp\left[-\left(\left(\dfrac{x}{\beta_1} \right)^{\alpha_1} + \left(\dfrac{y}{\beta_2} \right)^{\alpha_2} \right) \right]\times\\
%          \left(1+\rho\exp\left[-(a-1)\left(\left(\dfrac{x}{\beta_1} \right)^{\alpha_1}+ \left(\dfrac{y}{\beta_2} \right)^{\alpha_2}\right)\right]\right.\times\\
%          \left.\left(1-\exp\left[-\left(\dfrac{x}{\beta_1}\right)^{\alpha_1} \right] \right)^b \left(1-\exp\left[-\left(\dfrac{y}{\beta_2}\right)^{\alpha_2} \right]\right)^b\right)
%     \end{split}}
%  \end{split}
%\end{equation}
 \begin{equation}\label{hazardweibullfgm}
 \begin{split}
 h_2(x,y) = &\dfrac{\begin{split}
\dfrac{\alpha_1 \alpha_2}{\beta_1 \beta_2}A^{1-1/\alpha_1}B^{1-1/\alpha_2}\exp(-C)
    \left[1+\rho \left(1-\exp\left(-A\right)\right)^{(b-1)}
    \left(1-\exp\left(-B\right)\right)^{(b-1)}D\right]\end{split}}{\begin{split}
         \exp\left(-C \right) \left[1+\rho\exp\left(-(a-1)C\right)\left(1-\exp\left(-A \right) \right)^b \left(1-\exp\left(-B \right)\right)^b\right]
     \end{split}}
  \end{split}
\end{equation}
where $x>0$, $y>0$, $\alpha_1>0$, $\beta_1>0$, $\alpha_2>0$, $\beta_2>0$, $a\geq 1$, $b\geq 1$, $-1\leq \rho \leq 1$.

The hazard function for the  MBW distribution \eqref{hazardfun} can be expressed as 
\begin{equation}
    h(x,y)=\dfrac{f(x,y)}{R(x,y)}=w(x,y)h_1(x,y)+(1-w(x,y))h_2(x,y)
\end{equation}
where $w(x,y)=\dfrac{pR_1(x,y)}{R(x,y)}$ for all $x>0,\ y>0$.\\
Therefore, 
\begin{equation}\label{w}
 w(x,y) =\left\{
        \begin{array}{ll}
            \dfrac{p}{R(x,y)}, & \quad  x \leq x_{0},\  y \leq y_{0} \\
            \dfrac{p(x_0+d-x)}{dR(x,y)}, & \quad x_0\leq x \leq x_0+d,\  y \leq y_0\\
            \dfrac{p(y_0+d-y)}{dR(x,y)}, & \quad x\leq x_0 ,\  y_0 \leq y \leq y_0+d\\
             \dfrac{p(x_0+d-x)(y_0+d-y)}{d^2R(x,y)}, & \quad  x_{0} \leq x \leq x_{0}+d,\ y_{0} \leq y \leq y_{0}+d\\
             0, & \quad  x_0+d \leq x \ \text{or}\  y_0+d \leq y
        \end{array}
    \right.
\end{equation}
Thus, the expression for $R(x,y)$, $f(x,y)$ and $h(x,y)$ are given by
\begin{equation}\label{R}
 R(x,y) =\left\{\begin{array}{ll}
            p +qR_2(x,y), & \quad  x \leq x_{0},\ y \leq y_{0} \\
            \dfrac{p(x_0+d-x)}{d}+qR_2(x,y), & \quad x_0\leq x \leq x_0+d,\  y \leq y_0\\
            \dfrac{p(y_0+d-y)}{d}+qR_2(x,y), & \quad x\leq x_0 ,\  y_0 \leq y \leq y_0+d\\
             \dfrac{p(x_0+d-x)(y_0+d-y)}{d^2}+qR_2(x,y), & \quad  x_{0} \leq x \leq x_{0}+d,\ y_{0} \leq y \leq y_{0}+d\\
             qR_2(x,y), & \quad  x_0+d \leq x \ \text{or}\  y_0+d \leq y
        \end{array}
    \right.
\end{equation}
\begin{equation}\label{f}
 f(x,y) = \left\{
        \begin{array}{ll}
            \dfrac{p}{d^2}+qf_{XY}(x,y), & \quad x_{0} \leq x \leq x_{0}+d,\ y_{0} \leq y \leq y_{0}+d \\
            qf_{XY}(x,y), & \quad \text{Otherwise}
        \end{array}
    \right.
\end{equation}

\begin{equation}\label{h}
 h(x,y) =\left\{\begin{array}{ll}
           \dfrac{qf_{XY}(x,y)}{p +qR_2(x,y)}, & \quad  x \leq x_{0},\  y \leq y_{0}\\
            \dfrac{qdf_{XY}(x,y)}{p(x_0+d-x)+qdR_2(x,y)}, & \quad x_0\leq x \leq x_0+d,\  y \leq y_0\\
            \dfrac{qdf_{XY}(x,y)}{p(y_0+d-y)+qdR_2(x,y)}, & \quad x\leq x_0 ,\  y_0 \leq y \leq y_0+d\\
             \dfrac{p+qd^2f_{XY}(x,y)}{p(x_0+d-x)(y_0+d-y)+qd^2R_2(x,y)}, & \quad  x_{0} \leq x \leq x_{0}+d,\ y_{0} \leq y \leq y_{0}+d\\
             \dfrac{f_{XY}(x,y)}{R_2(x,y)}, & \quad  x_0+d \leq x \ \text{or}\  y_0+d \leq y
        \end{array}
    \right.
\end{equation}
%%$h(x,y)$ is continuous and differentiable $???????????????$%% 

	\section{Nearly instantaneous failure case $\left[(x_0,y_0) =(0,0)\right]$}\label{nearlyinstfailure}
\begin{figure}[H]
		\centering {\includegraphics[width=15cm]{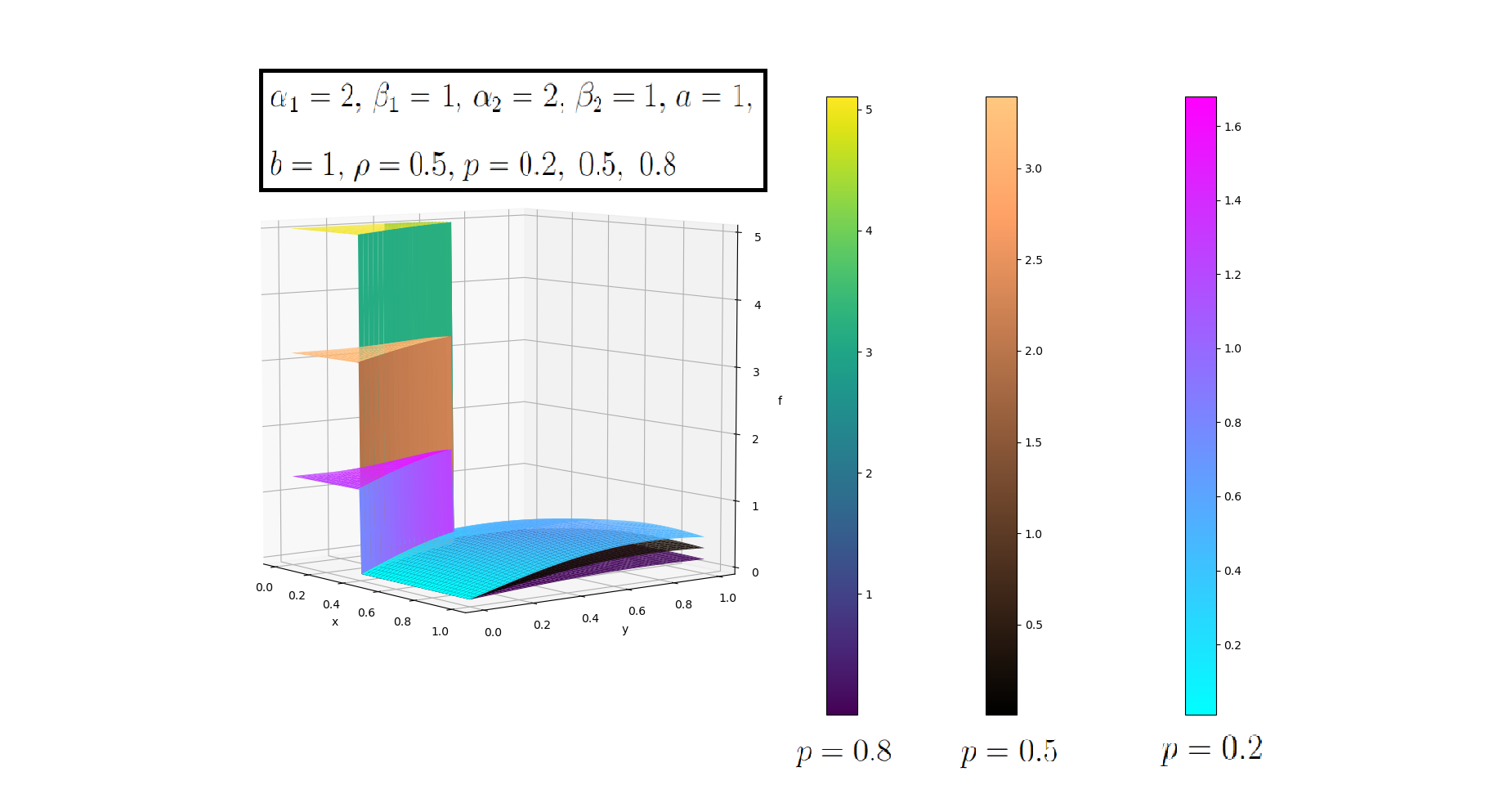}}
		\hspace{10em}
		\caption{Plot of MBW Density function with parameters: $\alpha_1=2$, $\beta_1=1$, $\alpha_2=2$, $\beta_2=1$, $a=1$, $b=1$, $\rho=0.5$, $d=0.4$, $p=0.2,\ 0.5, \  0.8$.}
		\label{fig: joint density for MBWD}
	\end{figure}

\begin{figure}[H]
		\centering {\includegraphics[width=15cm]{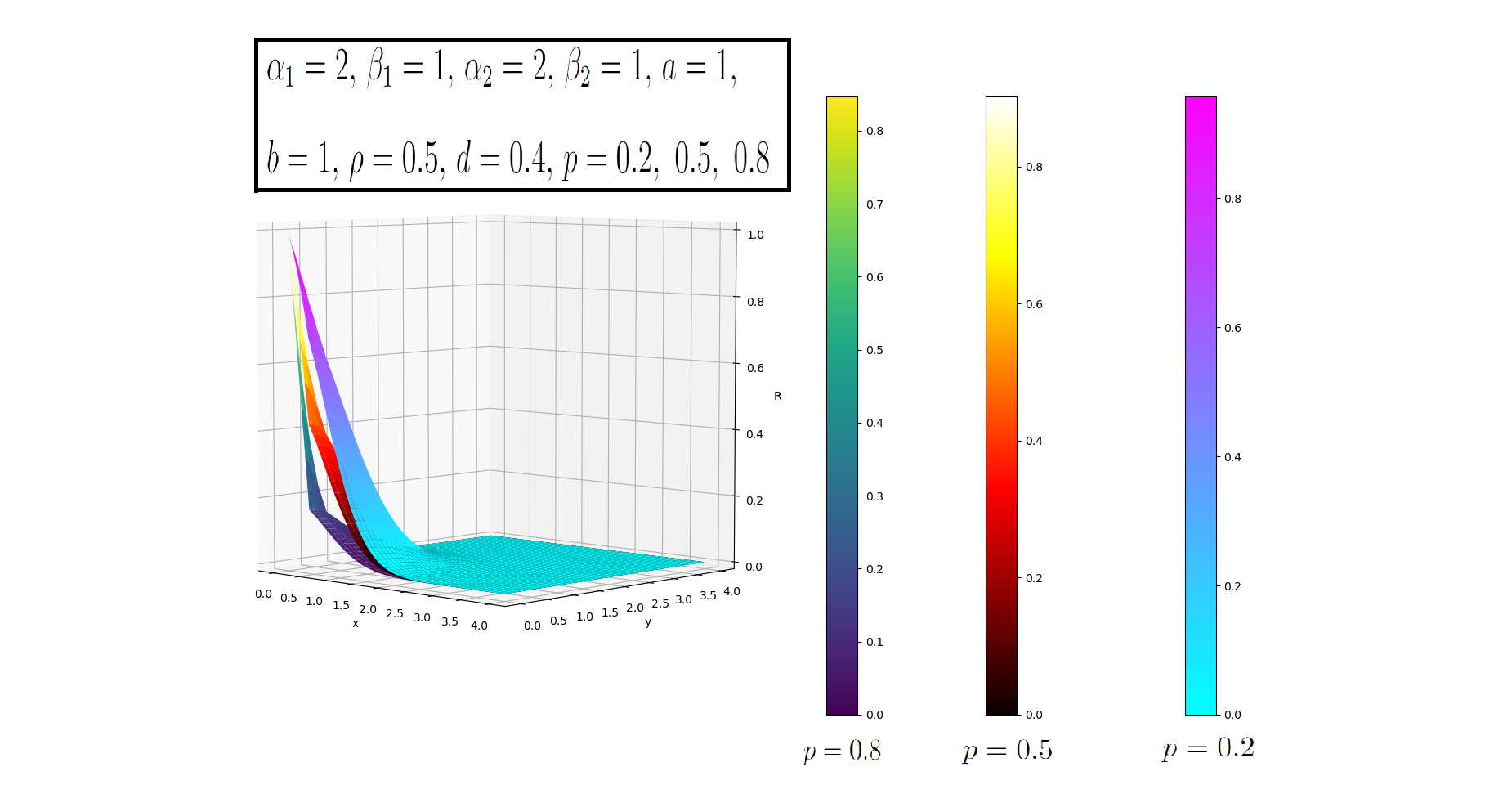}}
		\hspace{10em}
		\caption{Plot of survival function with parameters: $\alpha_1=2$, $\beta_1=1$, $\alpha_2=2$, $\beta_2=1$, $a=1$, $b=1$, $\rho=0.5$, $d=0.4$, $p=0.2,\ 0.5, \  0.8$.}
		\label{fig: survival funciton for MBWD}
\end{figure}

\begin{figure}[H]
		\centering {\includegraphics[width=15cm]{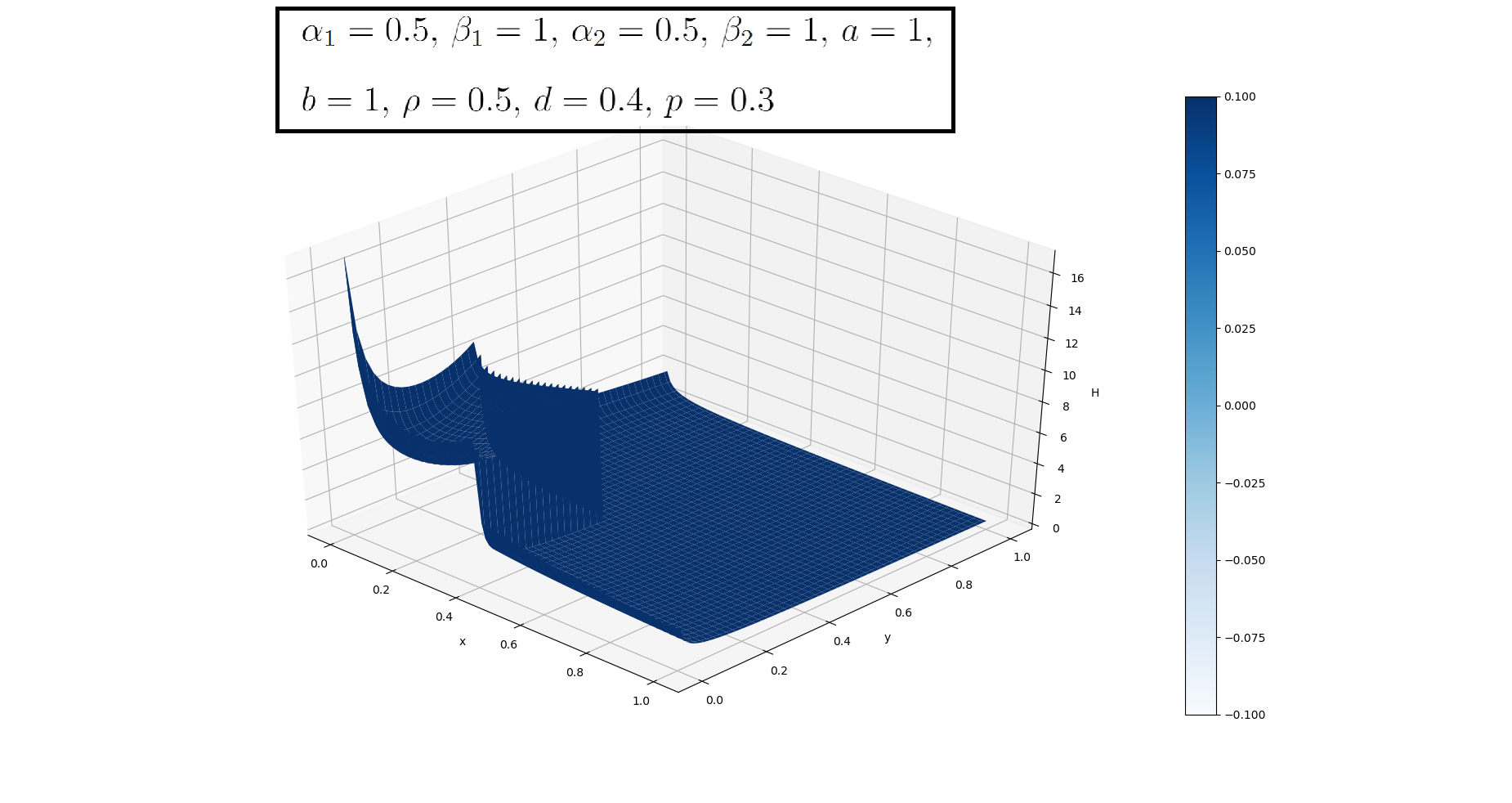}}
		\hspace{10em}
		\caption{Plot of hazard function with parameters: $\alpha_1=0.5$, $\beta_1=1$, $\alpha_2=0.5$, $\beta_2=1$, $a=1$, $b=1$, $\rho=0.5$, $d=0.4$, $p=0.3$.}
		\label{fig: hazard funciton_bathtub for MBWD}
	\end{figure}

\begin{figure}[H]
		\centering {\includegraphics[width=15cm]{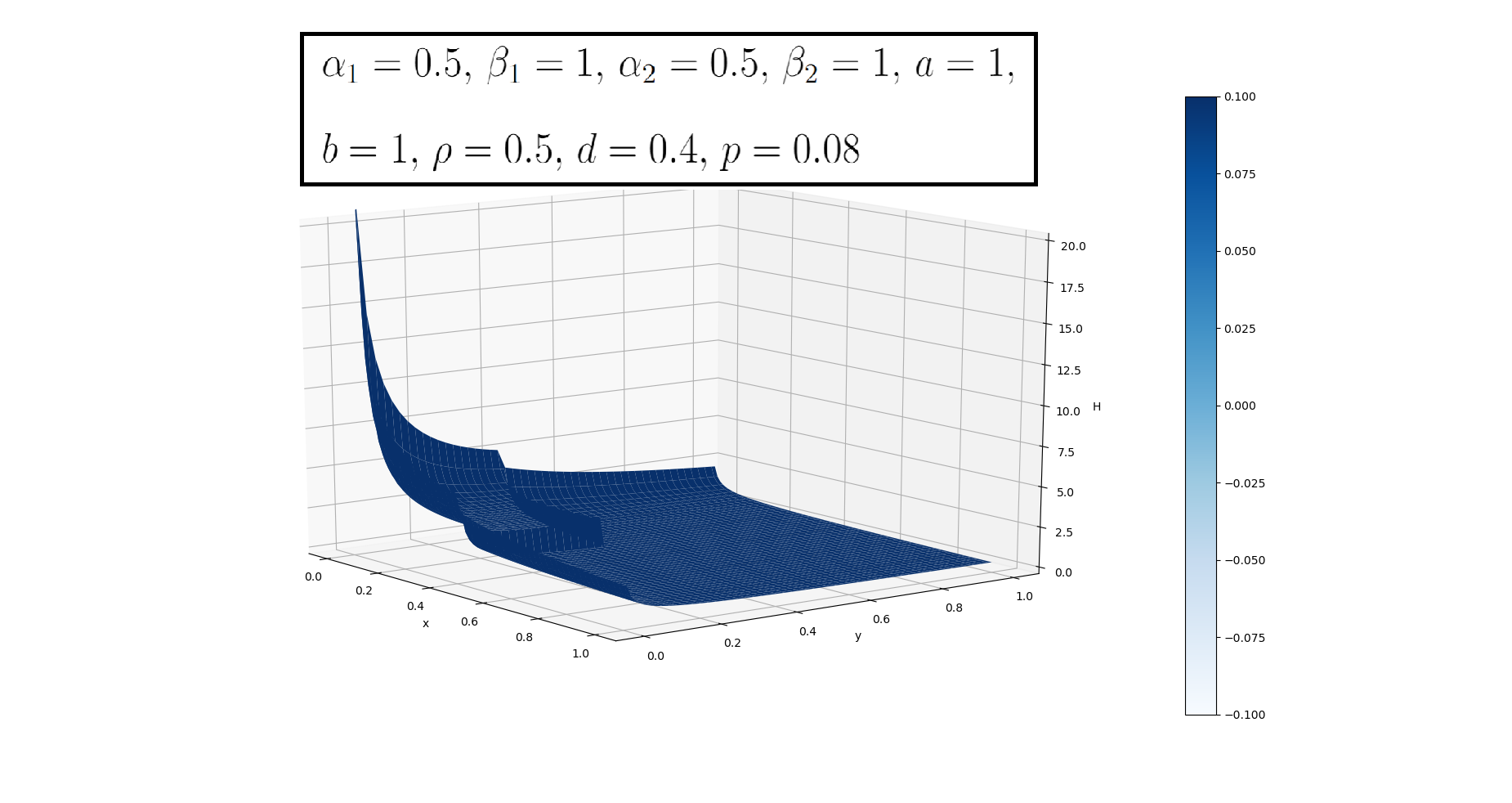}}
		\hspace{10em}
		\caption{Plot of hazard function with parameters: $\alpha_1=0.5$, $\beta_1=1$, $\alpha_2=0.5$, $\beta_2=1$, $a=1$, $b=1$, $\rho=0.5$, $d=0.4$, $p=0.08$.}
		\label{fig: hazard funciton_dec for MBWD}
	\end{figure}

\begin{figure}[H]
		\centering {\includegraphics[width=15cm]{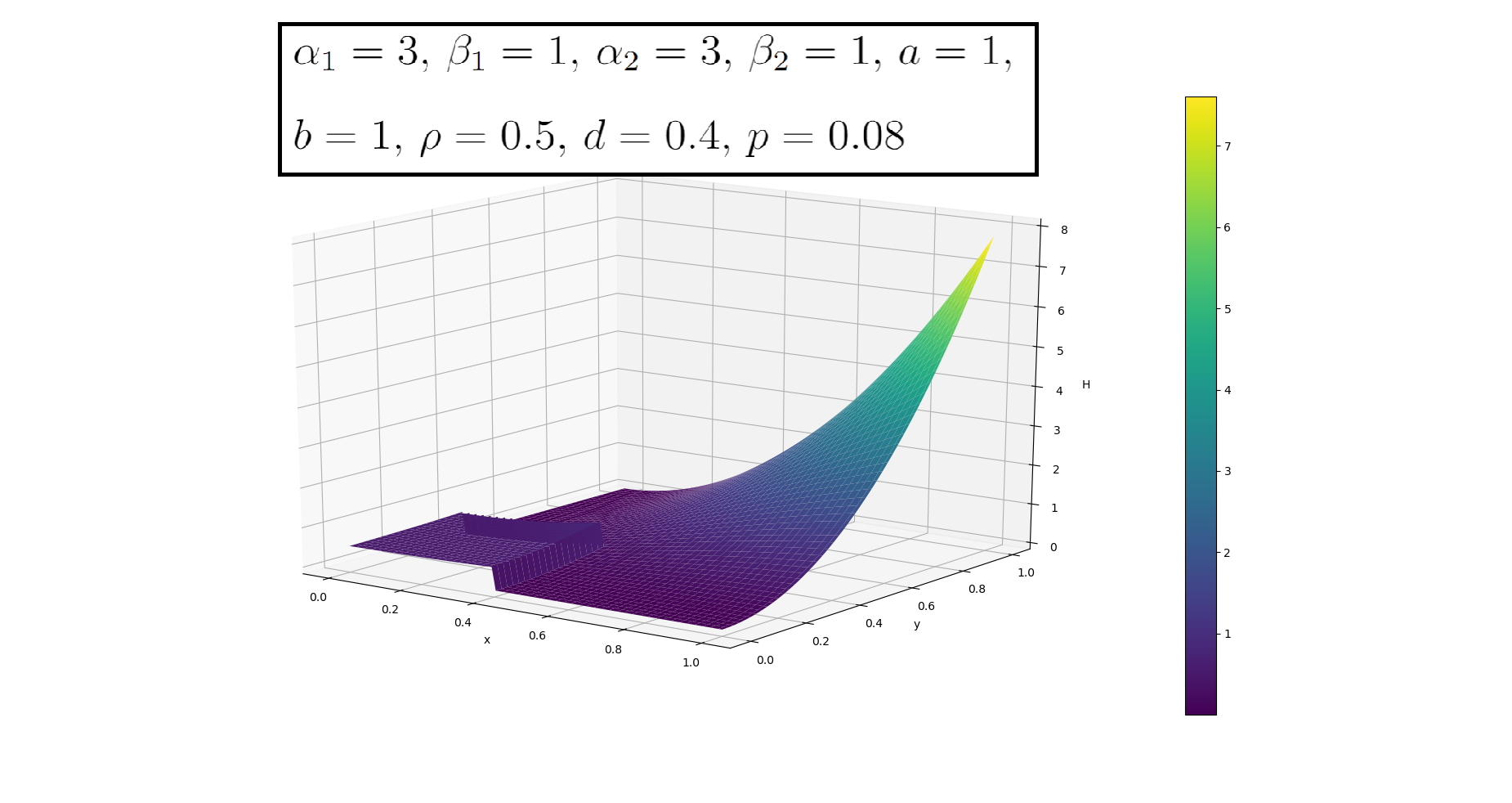}}
		\hspace{10em}
		\caption{Plot of hazard function with parameters: $\alpha_1=3$, $\beta_1=1$, $\alpha_2=3$, $\beta_2=1$, $a=1$, $b=1$, $\rho=0.5$, $d=0.4$, $p=0.08$.}
		\label{fig: hazard funciton_inc1 for MBWD}
	\end{figure}

\begin{figure}[H]
		\centering {\includegraphics[width=15cm]{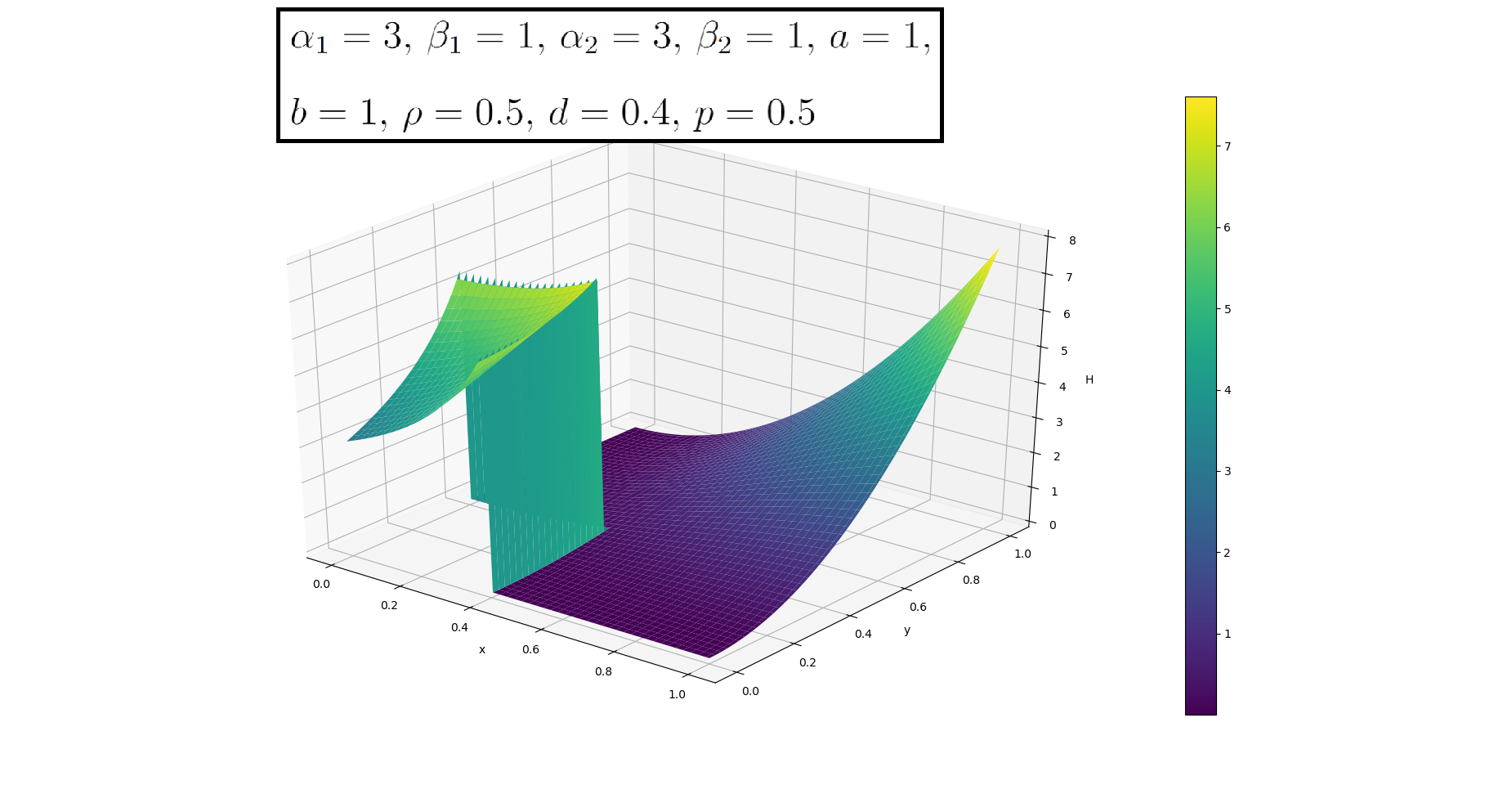}}
		\hspace{10em}
		\caption{Plot of hazard function with parameters: $\alpha_1=3$, $\beta_1=1$, $\alpha_2=3$, $\beta_2=1$, $a=1$, $b=1$, $\rho=0.5$, $d=0.4$, $p=0.5$.}
		\label{fig: hazard funciton_inc2 for MBWD}
	\end{figure}

Consider $(x_0,y_0)=(0,0)$ as a special case of the modified model \eqref{modifiedpdf}, which is called as nearly instantaneous failure model.
In this case the survival functions and hazard function for the bivariate uniform distribution are given by 
\begin{equation}\label{instantsurvivaldirac}
 R_1(x,y) =\left\{
        \begin{array}{ll}
             \dfrac{(d-x)(d-y)}{d^2}, & \quad  0 \leq x \leq d,\ 0 \leq y \leq d\\
             0, & \quad  d \leq x \ \text{or}\  d \leq y
        \end{array}
    \right.
\end{equation}
and 
\begin{equation}\label{instanthazarddirac}
 h_1(x,y) =\left\{
        \begin{array}{ll}
            \dfrac{1}{(d-x)(d-y)}, & \quad 0 \leq x \leq d,\ 0 \leq y \leq d\\
            \infty, & \quad  d \leq x \ \text{or}\  d \leq y.
        \end{array}
    \right.
\end{equation}
The survival functions and hazard function for the modified bivariate Weibull model with nearly instantaneous failure occurring uniformly over the rectangle $[0,d]\times[0,d]$ is given by
\begin{equation}\label{instantR}
 R(x,y) =\left\{\begin{array}{ll}
             \dfrac{p(d-x)(d-y)}{d^2}+qR_2(x,y), & \quad  0 \leq x \leq d,\ 0 \leq y \leq d\\
             qR_2(x,y), & \quad  d \leq x \ \text{or}\  d \leq y
        \end{array}
    \right.
\end{equation}
and 
\begin{equation}\label{instanth}
 h(x,y) =\left\{\begin{array}{ll}
             \dfrac{p+qd^2f_{XY}(x,y)}{p(d-x)(d-y)+qd^2R_2(x,y)}, & \quad  0 \leq x \leq d,\ 0 \leq y \leq d\\
             \dfrac{f_{XY}(x,y)}{R_2(x,y)}, & \quad  d \leq x \ \text{or}\  d \leq y
        \end{array}
    \right.
\end{equation}
Graphical plots are important to identify whether the models useful for specific data sets for which empirical plots are available. For the instantaneous failure case, the graph of the MBW density function, survival function and hazard rate function are given for specific parameter values in the following way.
\begin{itemize}
    \item \textbf{Density function}: The plot of the density functions for the parameter values $\alpha_1=2$, $\beta_1=1$, $\alpha_2=2$, $\beta_2=1$, $a=1$, $b=1$, $\rho=0.5$, $d=0.4$ and $p=0.2,\ 0.5,\ 0.8$ are given in Figure \ref{fig: joint density for MBWD}.
    \item \textbf{Survival function}: Corresponding to the density functions given in Figure 1, the plot of the survival functions are given in Figure \ref{fig: survival funciton for MBWD}.
    \item \textbf{Hazard function}: The plot of the hazard function  are given in the Figures \ref{fig: hazard funciton_bathtub for MBWD}, \ref{fig: hazard funciton_dec for MBWD}, \ref{fig: hazard funciton_inc1 for MBWD} and \ref{fig: hazard funciton_inc2 for MBWD} for different parameters. We can observe that the graph is similar to the Weibull hazard function outside the rectangle $[0,d]\times [0,d]$. Inside the rectangle, the behavior of the hazard function has more than one direction where it takes increasing, decreasing, and bathtub shape.
    
\end{itemize}

	\section{Estimation parameters}\label{estimation}
We estimate the unknown parameters of the proposed model using the MLE combined with the DBSCAN algorithm as follows. 
The joint PDF of the modified bivariate Weibull distribution allowing early and instantaneous failures is given by
\begin{equation}\label{nf}
 f(x,y) = \left\{
        \begin{array}{ll}
            \dfrac{p}{d^2}+qf_{XY}(x,y), & \quad 0 \leq x \leq d,\ 0 \leq y \leq d \\
            qf_{XY}(x,y), & \quad \text{Otherwise,}
        \end{array}
    \right.
\end{equation}
where $f_{XY}$ is the joint PDF of bivariate Weibull distribution given in \eqref{bivariateweibullpdfuc}. 
%Here we consider two marginal two parameters Weibull distribution with parameters $\alpha_1,\beta_1; \alpha_2, \beta_2 $ and the Gaussian copula with correlation coefficient $\rho$ to get the join PDF of the modified distribution.%
The unknown parameters of the model are given by $\Theta=\{\alpha_1,\beta_1, \alpha_2, \beta_2,\bm{\rho}, d, p \}$ with $\bm{\rho}$  representing copula parameters. To estimate the parameters $\Theta$, we combine the MLE technique and the DBSCAN clustering method in the following way.\\
Let $(x_i,y_i),\ i=1,2, \cdots, n$ be the $n$ sample points from the population with PDF given by \eqref{nf}. Then, the likelihood function is 
\begin{equation}\label{likelihood}
 L(\Theta|\textbf{x,y})= \prod_{0 \leq x_i \leq d,\ 0 \leq y_i \leq d}\left[\dfrac{p}{d^2}+qf_{XY}(x_i,y_i)\right]\times \prod_{x_i>d\ or\ y_i>d}qf_{XY}(x_i,y_i).
\end{equation}
and the corresponding log-likelihood function is 
\begin{equation}\label{loglikelihood}
\begin{split}
 l(\Theta|\textbf{x,y})&=log[ L(\Theta|\textbf{x,y})]\\
 &=\sum_{0 \leq x_i \leq d,\ 0 \leq y_i \leq d}log\left[\dfrac{p}{d^2}+qf_{XY}(x_i,y_i)\right]+\sum_{x_i>d\ or\ y_i>d}log\left[qf_{XY}(x_i,y_i)\right].
\end{split}
\end{equation}

We usually maximize the log-likelihood function to find the estimates of unknown parameters of the models by differentiating the likelihood or log-likelihood function with respect to the parameters and solving the equations simultaneously. However, we do not get a definite solution for d by the said approach due to the particular nature of the likelihood function concerning the corresponding variable. 
Now, we see that the log-likelihood function decreases as the value of d increases. So, the maximum log-likelihood function value occurred at the minimum possible value of d. Now, the contribution of d in the likelihood function is non-zero when d satisfies the following inequalities: $0<x_i<d$ and $0<y_i<d$ where $x_i$ and $y_i$ are independently generated from uniform distribution in [0,d]. So, we classify the data using the DBSCAN clustering method given by \cite{ester1996density}. The DBSCAN parameters are estimated using the methodologies  provided in \cite{schubert2017dbscan} and  \cite{rahmah2016determination}.  Then two clusters are formed; one is C1, generated from the bivariate uniform distribution, and another C2 from bivariate Weibull distribution. Now, we find the maximum likelihood estimate of d given by,  
\begin{equation}\label{destimate}
    \hat{d}=\max\{\max\{x_i,y_i\}|(x_i,y_i)\in C1\}
\end{equation}
Then, we use the usual approach of finding the MLEs of other parameters by maximizing the log-likelihood function using the estimated value $\hat{d}$. In the numerical examples given in Section \ref{numaricalexample}, we use R-packages ``optim" and ``dbscan" for estimating the model parameters using the proposed method.

  \section{Algorithm for Data Simulation}\label{algorithm}
In this section, we discuss an algorithm for simulating data (x,y) from the MBW distribution given in \eqref{modifiedpdf}. The steps are as follows:
\begin{enumerate}
    \item Generates a value z from  binomial distribution B(1,p).
    \item If $z=1$, then Generate x and y independently from the uniform distribution $U(0,d)$.
    %then generates (x,y) from the bivariate uniform distribution, given in \eqref{bivariatedirac}, in the following ways; Generate x and y independently from the uniform distribution $U(0,d)$.
    \item If $z=0$, then generates (x,y) from the bivariate Weibull distribution given in \eqref{bivariateweibullpdfuc}, in the following way;
    \begin{enumerate}
        \item Generate $u$ and $t$ independently from uniform $(0,1)$.
        \item Let $v=C^{-1}_u(t)$, where $C^{-1}_u$ is the quasi-inverse of $C_u$. Here, $C_u$ is defined as,  $C_u:=P(V\leq v|U=u)=\dfrac{\partial C(u,v)}{\partial u}$, where C is a  bivariate copula.\\
          In the case of the GFGM copula, we can find $v$ by solving the following equation using the Newton Raphson iteration method.
    \begin{equation*}
        v+\rho b u^{b-1}v^b(1-u)^a(1-v)^a-\rho a u^bv^b(1-u)^{a-1}(1-v)^a-t=0
    \end{equation*}
        \item Set $x= F_1^{-1}(u)$ and $y=F_2^{-1}(v)$, where $F_1$ and $F_2$ are the CDF of univariate Weibull distribution with parameters  $(\alpha_1, \beta_1 )$ and $(\alpha_2, \beta_2 )$ respectively.
    \end{enumerate}
\end{enumerate}
To simulate n-data points using this algorithm we repeat these three steps n times.

	\section{Numerical Example}\label{numaricalexample}
 Here, we provide numerical examples using simulated data to illustrate and test the proposed methodologies and apply them to real data.
 %with considering the Gaussian copula.
 %%and also apply the model distribution to a real data. %%
  \subsection{Simulation Studies} \label{simulation}
	We simulate  data $(x_i,y_i)$, for $i=1,2,3,\cdots n$, using the algorithm proposed in Section \ref{algorithm} from modified bivariate Weibull distribution \eqref{modifiedpdf} with parameter values assumed as  $\Theta=(\alpha_1=4,\  \beta_1=1.5,  \alpha_2=3.5,\  \beta_2=5,\ \rho=0.6,\ d=0.1,\ p=0.3)$. Three different samples sizes $n=100,\ 200,$ and $300$ are considered for our simulation studies. We fit the simulated data using the proposed modified bivariate Weibull distribution \eqref{modifiedpdf} assuming the parameter vector $\Theta$ as unknown. 
	
%%	Usually, we assume that the multivariate count data of life follows bivariate Weibull distribution. But we may get a better result if we consider the mixture of the generalized bivariate Dirac delta function and the bivariate two-parameter Weibull distribution. The bivariate Weibull distribution is obtained using FGM copula having margins univariate two parameters Weibull distribution.\\%%
	
The unknown parameter vector $\Theta=\{\alpha_1,\beta_1, \alpha_2, \beta_2, \rho, d, p \}$ is estimated using the proposed methodologies given in Section \ref{estimation}. We first find the estimate $\hat{d}$ of  $d$ using the DBSCAN clustering algorithm and the statistic given in \eqref{destimate}. For the sample sizes $n=100, 200$, and $300$, the estimated parameter values of DBSCAN $[Minpts=4, eps=0.45]$, $[Minpts=4, eps=0.35]$, and $[Minpts=4, eps=0.25]$ respectively. Then, the log-likelihood function is maximized to find the MLEs of the other model parameters using the value of $d$ as $\hat{d}$. For each sample size $n\in\{100,200,300\}$, we simulate 2000 data sets and estimate the model parameters by the above methods. Then, we find the sample mean, Bias, Mean Square Error (MSE), bootstrap standard error (BSE), and 95\% bootstrap confidence interval (BCI) using 2000 MLEs of each parameter for comparing the estimated values with the true parameter values of the model. A 95\% confidence interval is also computed using the asymptotic results of MLE, and the corresponding coverage probabilities are also found. The coverage probability is considered as a percentage of times that the confidence interval includes the respective true parameter values. To calculate the confidence interval of $d$ for each sample data, we consider the univariate data $\{x_i|(x_i,y_i)\in C_1\}\cup \{y_i|(x_i,y_i)\in C_1\}$ of size $2n$ which follows uniform distribution on $[0,d]$, where $C_1$ is the cluster near origin. The MSE and the Bias are calculated using the formula 
	\begin{equation}
	\begin{split}
	     Bias(\hat{\theta})=\dfrac{1}{N}\sum_{i=1}^{N}\hat{\theta}^{(i)}-\theta\\
	     MSE(\hat{\theta})=\dfrac{1}{N}\sum_{i=1}^{N}\left(\hat{\theta}^{(i)}-\theta\right)^2 
	\end{split}
	\end{equation}
where $\hat{\theta}$ denotes the estimator of the parameter $\theta$ and $\hat{\theta}^{(i)}$ denotes the estimate of $\theta$ for $i$th data set, where $N$ is the number of data generated with sample size $n$.

The sample mean, coverage probability, Bootstrap confidence interval (BCI), MSE and Bias using 2000 estimated parameter values with sample sizes $n=100, 200, 300$ are given in Table \ref{Table-100}, Table \ref{Table-200} and Table \ref{Table-300} respectively. We see that the Bias and MSE  are small, i.e., the estimated parameter values by the proposed methodologies are close to the true parameter values, and the coverage probabilities are approximately the same as the confidence coefficients for each scenario with varying sample sizes. The bootstrap confidence intervals provide expected results for estimating model parameters using the proposed methodologies.

	 \begin{table}[h!]
	\centering
	\begin{tabular}{||c c c c c c c||} 
		\hline
		Parameters & Sample Mean & CP & BSE & BCI & MSE & Bias\\ [0.5ex] 
		\hline\hline
		$\alpha_{1}=4$ &  4.0823 & 0.95 & 0.4360 & (3.4055\ 4.9091) &  0.1967  & 0.0823\\ 
		\hline
		$\beta_{1}=1.5$ & 1.4982 & 0.94 & 0.0581 & (1.4070\ 1.5939)  &  0.0033  & -0.0017 \\
		\hline
		$\alpha_{2}=3.5$ & 3.5628 & 0.94 & 0.3936 & (2.9375\ 4.3292) & 0.1588 & 0.0628 \\ 
		\hline
        $\beta_{2}=5$ & 4.9818 & 0.94 & 0.2711 & (4.6446\  5.3622) & 0.0738 & -0.0181 \\
		\hline
		$\rho=0.6$ & 0.5970 & 0.93 & 0.0789 & (0.4342\  0.7394) & 0.0062 & -0.0029 \\ 
		\hline
		$d=0.1$ & 0.1005 & 0.94 & 0.0284 & (0.0934\  0.0999) & 0.0008 & 0.0005 \\
		\hline
		$p=0.3$ & 0.2969 &  0.95 & 0.0507 &  (0.2099\  0.3899) &  0.0025 & -0.0030\\ 
		\hline\hline
	\end{tabular}
	\caption{Sample mean of estimated parameters, coverage probability(CP), 95\% bootstrap confidence interval(BCI), MSE and Bias for 100 data points}
	\label{Table-100}
\end{table}

%estimates 4.006807 1.49357 3.504285 4.95358 0.5981616 0.2988985 0.1033949
%CP 94.15 93.95 94.25 93.75 92.9 92.25 93.25
% bse  0.4399697 0.08520202 0.3918012 0.4224052 0.06792517 0.0516276 0.03833179
%BCI [1,] 3.49563986 4.6019809
%[2,] 1.42771746 1.5654279
%[3,] 3.06699064 4.0266522
%[4,] 4.71577913 5.2445313
%[5,] 0.48515092 0.6995568
%[6,] 0.23499622 0.3700010
%[7,] 0.09690768 0.0999908
% MSE 0.1935229 0.007297096 0.1534498 0.1804918 0.004614901 0.002665289 0.001480117
% bias 0.006807416 -0.006429704 0.004284691 -0.04642049 -0.00183841 -0.001101503 0.003394857

% edit for ss200
	 \begin{table}[h!]
	\centering
	\begin{tabular}{||c c c c c c c||} 
		\hline
		Parameters & Sample Mean & CP & BSE & BCI & MSE & Bias\\ [0.5ex] 
		\hline\hline
		$\alpha_{1}=4$ & 4.0068 & 0.94 & 0.4399 & (3.4956\  4.6019) &  0.1935  & 0.0068\\ 
		\hline
		$\beta_{1}=1.5$ & 1.4935 & 0.94 &  0.0852 & (1.4277\  1.5654)  &  0.0072  & -0.0064 \\
		\hline
		$\alpha_{2}=3.5$ & 3.5042  & 0.94 & 0.3918 & (3.0669\  4.0266) & 0.1534 & 0.0042 \\ 
		\hline
        $\beta_{2}=5$ & 4.9535 & 0.94 & 0.4224 & (4.7157\  5.2445) & 0.1804 & -0.0464 \\
		\hline
		$\rho=0.6$ & 0.5981 & 0.93 & 0.0679 & (0.4851\  0.6995) & 0.0046 & -0.0018 \\ 
		\hline
		$d=0.1$ & 0.1033 & 0.92 & 0.0383  & (0.0969\  0.0999) & 0.0014 & 0.0033\\
		\hline
		$p=0.3$ & 0.2988 &  0.93 & 0.0516 &  (0.2349\  0.3700) &  0.0026 & -0.0011\\ 
		\hline\hline
	\end{tabular}
	\caption{Sample mean of estimated parameters, coverage probability(CP), 95\% bootstrap confidence interval(BCI), MSE and Bias for 200 data points}
	\label{Table-200}
\end{table}

	 \begin{table}[h!]
	\centering
	\begin{tabular}{||c c c c c c c||} 
		\hline
		Parameters & Sample Mean & CP & BSE & BCI & MSE & Bias\\ [0.5ex] 
		\hline\hline
		$\alpha_{1}=4$ &  4.0225 & 0.93 & 0.4264 & (3.3936\  4.7940) &  0.1822  & 0.0225\\ 
		\hline
		$\beta_{1}=1.5$ & 1.4939 & 0.94 &  0.0660 & (1.4055\  1.5787)  &  0.0043  & -0.0060 \\
		\hline
		$\alpha_{2}=3.5$ & 3.5349 & 0.94 & 0.3782 & (2.9488\  4.2028) & 0.1442 & 0.0349 \\ 
		\hline
        $\beta_{2}=5$ & 4.9768 & 0.94 & 0.2719 & (4.6402\  5.3101) & 0.0744 & -0.0231 \\
		\hline
		$\rho=0.6$ & 0.6010 & 0.94 & 0.0722 & (0.4573\  0.7296) & 0.0052 & 0.0010 \\ 
		\hline
		$d=0.1$ & 0.1025 & 0.95 & 0.0360  & (0.0946\  0.0999) & 0.0013 & 0.0025 \\
		\hline
		$p=0.3$ & 0.2984 &  0.94 & 0.0451 &  (0.2199\  0.3800) &  0.0020 & -0.0015\\ 
		\hline\hline
	\end{tabular}
	\caption{Sample mean of estimated parameters, coverage probability(CP), 95\% bootstrap confidence interval(BCI), MSE and Bias for 300 data points}
	\label{Table-300}
\end{table}

%	 \begin{table}[h!]
%	\centering
%	\begin{tabular}{||c c c c c c c||} 
%		\hline
%		Parameters & Estimates & CP & BSE & BCI & MSE & Bias\\ [0.5ex] 
%		\hline\hline
%		$\alpha_{1}=4$ &  3.9350 & 0.93 & 0.0092 & (1.6071\  4.3500) &  0.2634  & -0.0649\\ 
%		\hline
%		$\beta_{1}=1.5$ & 1.4866 & 0.93 &  0.0014 & (1.4435\  1.5417)  &  0.0109  & -0.0133 \\
%		\hline
%		$\alpha_{2}=3.5$ & 3.4395 & 0.92 & 0.0083 & (1.4919\  3.8146) & 0.2117 & -0.0604 \\ 
%		\hline
%        $\beta_{2}=5$ & 4.9252 & 0.92 & 0.0063 & (3.2876\  5.1606) & 0.2350 & -0.0747 \\
%		\hline
%		$\rho=0.6$ & 0.6034 & 0.93 & 0.0016 & (0.5279\  0.6789) & 0.0041 & 0.0034 \\ 
%		\hline
%		$d=0.1$ & 0.1221 & 0.91 & 0.0008  & (0.09874\  0.4054) & 0.0633 & 0.0221\\
%		\hline
%		$p=0.3$ & 0.2958 &  0.93 & 0.0010 &  (0.2439\  0.3419) &  0.0031 & -0.0041\\ 
%		\hline\hline
%	\end{tabular}
%	\caption{Parameter estimates, coverage probability(CP), 95\% bootstrap confidence interval(BCI), MSE and Bias for 500 data points}
%	\label{Table-500}
%\end{table}

\subsection{Application to Vannman wood data}\label{real_data}
An experiment was done by Vannman (1991) to compare two chemical processes for drying wooden boards. Two different chemical processes are used to dry a batch of wooden boards under the same climate conditions. In the experiment, two chemical processes are compared, and damage to the board is measured as the percentage of the checking area given in \cite{vanman91}. As not all the boards are checked, the sample contains some zero observation. This observation corresponds to instantaneous failure or early failure. The reproduced data set of the experiment on two batches of 36 boards by using two different schedules are given in Table \ref{realdata}.  

%Schedule 1: $x_i=0$, $i=0,1,\cdots, 13$ and the others positive observations are 0.08, 0.32, 0.38, 0.46, 0.71, 0.82, 1.15, 1.23, 1.40, 3.00, 3.23, 4.03, 4.20, 5.04, 5.36, 6.12, 6.79, 7.90, 8.27, 8.62, 9.50, 10.15, 10.58, and 17.49. \\

%Schedule 2:  $x_i=0$, $i=0,1,\cdots, 17$ and the others positive observations are 0.02, 0.02, 0.02, 0.04, 0.09, 0.23, 0.26, 0.37, 0.93, 0.94, 1.02, 2.23, 2.79, 3.93, 4.47, 5.12, 5.19, 5.39, 6.83, and 8.22. 
\begin{table}[h!]
    \centering
\begin{tabular}{||c| c c ||c| c c||} \hline
Sl. no.& schedule 1 $(y_{1})$ & schedule 2 $(y_{2})$ & Sl. no & schedule 1 $(y_{1})$ & schedule 2 $(y_{2})$ \\ [0.5ex] 
 \hline\hline
 1 & 0 & 0 & 19 & 0.82  & 0,02 \\ 
 \hline
 2 & 0 & 0 & 20 & 1.15  & 0.02 \\ 
 \hline
 3 & 0 & 0 & 21 & 1.23& 0.04 \\ 
 \hline
 4 & 0 & 0 & 22 & 1.40  & 0.09 \\ 
 \hline
  5 & 0 & 0 & 23 & 3.00 & 0.23 \\ 
 \hline
  6 & 0 & 0 & 24 & 3.23  & 0.26 \\ 
 \hline
 7 & 0 & 0 & 25 & 4.03 & 0.37 \\ 
 \hline
 8 & 0 & 0 & 26 & 4.20  & 0.93 \\ 
 \hline
 9 & 0 & 0 & 27 & 5.04 & 0.94 \\ 
 \hline
 10 & 0 & 0 & 28 & 5.36  & 1.02 \\ 
 \hline
 11 & 0 & 0 & 29 & 6.12 & 2.23 \\ 
 \hline
 12 & 0 & 0 & 30  & 6.79  & 2.79 \\ 
 \hline
 13 & 0 & 0 & 31 & 7.90  & 3.93 \\ 
 \hline
 14 & 0.08  & 0 & 32 & 8.27  & 4.47 \\ 
 \hline
 15 & 0.32 & 0 & 33 & 8.62  & 5.12 \\ 
 \hline
 16 & 0.38  & 0 & 34 & 9.50  & 5.19 \\ 
 \hline
 17 & 0.46  & 0 & 35 & 10.15  & 5.39 \\ 
 \hline
 18 & 0.71 & 0.02 & 36 & 10.58 & 6.83 \\
 \hline
\end{tabular}
\caption{Vannman experiment data(1991) }
\label{realdata}
\end{table}

We consider three models, M1, M2, and M3, to fit the data for comparing the proposed model M3 with existing other models M1 and M2, frequently used in the literature. Model M1 assumes independent bivariate responses, and Model M2 considers dependent structure using a copula, ignoring the presence of instantaneous and early failure in the data. As the data contains many $(0,0)$ observations, it will be realistic to fix the shape parameter value $\alpha=1$ in the two parameters Weibull distribution, i.e., exponential distribution, otherwise for $\alpha \neq 1$, the log-likelihood value is not finite for model M1 and M2. For the same reason, Gaussian copula, t-copula, and Archimedean copulas are not applicable for the dependence structure in model M2, so we used the FGM copula with the parameter values  $a=1,\ b=1$. However, the proposed model M3 can take two-parameter Weibull distributions as their marginals and any copulas for the dependence structure. Here, we fit the data using  Gaussian, t-copula, Archimedean copula, and FGM copula, and the minimum AIC value selects the FGM copula for Model M3. 
 
The unknown parameters of Models M1 and M2 are estimated using the MLE, and the proposed methodologies given in Section \ref{estimation} are applied to estimate the unknown parameters of Model M3. The parameter $d$ in Model M3 is found using the estimated DBSCAN parameter values as $Minpts=4$, and $eps=1.6$. 
The estimates, likelihood values, and Akaike Information Criterion (AIC) values for Models M1, M2, and M3 are given in Table \ref{Table-M1}, Table \ref{Table-M2}, and Table \ref{Table-M3}. The AIC values of Models M1, M2, and M3 are 228.4698, 195.8067, and 165.2824, respectively, indicating that Model M3 is the best-fitted model to the data with the minimum AIC values as a criterion. We also observed that Model M3 provides a significantly better fit than Models M1 and M2 by testing the hypotheses using the deviance statistic, rejecting the Null hypothesis with P-values less than 0.0001. Hence, researchers may find a better performance by using the proposed model to fit bivariate continuous responses in the presence of instantaneous and early failure in the data. 
% SE 0.4599071 0.1846755
% p-val 1.973136e-09 1.973011e-09
	 \begin{table}[h!]
	\centering
	\begin{tabular}{||c c c c|} 
		\hline
		Parameters & Estimates & SE & p-value \\ [0.5ex] 
		\hline\hline
		$\beta_{1}$ & 2.759445 & 0.4599071 & $<0.0001$\\
		\hline
        $\beta_{2}$ & 1.108056 & 0.1846755 & $<0.0001$\\
		\hline \hline
		Likelihood value & -112.2349 & &\\
		\hline
		AIC & 228.4698 & & \\ 
		\hline\hline
	\end{tabular}
	\caption{
Parameter estimates, likelihood values, and AIC from model M1 for real data }
	\label{Table-M1}
\end{table}

%se2
%[1] 0.45868422 0.16424681 0.00000002
%pvalue2
%[1] 2.849280e-10 2.576438e-10 0.000000e+00 
	 \begin{table}[h!]
	\centering
	\begin{tabular}{||c c c c|} 
		\hline
		Parameters & Estimates & SE & p-value \\ [0.5ex] 
		\hline\hline
		$\beta_{1}$ & 2.8934625 &  0.4586 & $<0.0001$   \\
		\hline
        $\beta_{2}$ & 1.0333664 &  0.1642 & $<0.0001$ \\
        \hline
        $\rho$ & 0.9994815 & 0.00000002 & $<0.0001$ \\
		\hline \hline
		Likelihood value & -94.90337 &  & \\
		\hline
		AIC & 195.8067 & & \\ 
		\hline\hline
	\end{tabular}
	\caption{Parameter estimates, likelihood values, and AIC from model M2 for real data }
	\label{Table-M2}
\end{table}

%se3
% [1] 0.80926004 0.14057986 0.34929059 0.00000002 0.01645766 0.08475749
% > pvalue3
% [1] 1.034958e-05 0.000000e+00 5.308312e-05 0.000000e+00 0.000000e+00 2.938768e-12
	 \begin{table}[h!]
	\centering
	\begin{tabular}{||c c c c ||} 
		\hline
		Parameters & Estimates & SE & p-value \\ [0.5ex] 
		\hline\hline
		$\alpha_{1}$ & 2.691784 & 0.8092 & $<0.0001$ \\ 
		\hline
		$\beta_{1}$ & 7.739789 & 0.1405 & $<0.0001$\\
		\hline
		$\alpha_{2}$ & 1.000698 & 0.3492 &  $<0.0001$ \\ 
		\hline
        $\beta_{2}$ & 3.309285 & 0.00000002 & $<0.0001$ \\
		\hline
		$\rho$ & 0.9849288 & 0.0164 & $<0.0001$  \\ 
		\hline
		$d$ &  1.4 & 0.0017 & $<0.0001$  \\
		\hline
		$p$ & 0.5784499 &  0.0847 &  $<0.0001$  \\ 
		\hline \hline
		Likelihood value & -75.6412 & &  \\
		\hline
		AIC & 165.2824 &  & \\ 
		\hline\hline
	\end{tabular}
	\caption{Parameter estimates, likelihood values, and AIC from model M3 for real data }
	\label{Table-M3}
\end{table}

	\section{Conclusion}\label{conclusion}

We proposed a modified bivariate Weibull distribution for modeling data allowing early and instantaneous failure observations. The bivariate Weibull distribution is obtained using a copula, assuming the marginals are distributed as two parameters Weibull distribution. The survival and hazard functions are provided, and the explicit forms of the functions are given in the particular case of GFGM copula. We used MLE combined with DBSCAN clustering algorithm to estimate the unknown model parameters. Numerical examples are shown to illustrate and test the proposed method using simulated data. The coverage probability, bootstrap standard error, bootstrap confidence intervals, mean square error, and bias are calculated to test the performance of the proposed methodologies. The proposed model is applied to real data and compared with existing models in the literature. We observed that the modified Weibull distribution outperforms with respect to the AIC as a criterion and provides a significantly better fit by the chi-square test than other existing models in the literature.   

We have used the MLE combined with the DBSCAN clustering algorithm to estimate the parameters. Researchers may use different estimation methods, such as the Expectation-Maximization algorithm, Bayesian techniques, etc., to estimate the model's unknown parameters. One of the future directions may be to apply and compare different estimation techniques to have an optimum performance of the model.

%Our next goal is to apply the proposed model to bivariate real data and study the effect of different assumed copula for predicting responses using the fitted model.

%We proposed a regression model for multivariate multiple inflated negative binomial count responses. We illustrated the proposed method for modeling doubly inflated count responses using simulated data in numerical examples. Our next goal is to test the performance of the proposed models with various existing models using Poisson and negative binomial distributions, which did not consider the multiple inflated cells in the data. We will also study the effects of different assumed copula for predicting responses using the fitted model and propose a regression method robust to the misspecification of the form of the copula.

	%
	
	%
	%\section{Title of the section}
	%Text of the first section. \citet{Fahrmeiret13} is a~direct reference to a~book with more than two authors.
	%\citet{Gomezet09} is a~direct reference to a~journal article with more than two authors.
	%Many papers were published in \emph{Statistical Modelling} \citep[see, e.g.,][]{Kneib13, KomarekLesaffre06, Liet07, Waldmannet13}.
	%Sometimes we also need to reference a~book chapter \citep{Lesaffreet09}.
	%
	%\subsection{Title of the subsection}
	%
	%\subsubsection{Title of the subsubsection}

	%%% Acknowledgements (if any)
	%%% ------------------------------------------
	
	\newpage
	
%	\section*{Computer programs used for computations}
%	All the computations in the numerical examples to illustrate the proposed methodologies are performed using a language and environment for statistical computing called R \citep{R2020}.

%	\section*{Acknowledgements}
%	This work was supported by the Science and Engineering Research Board, Govt. of India (SRG/2019/001226).
	
%	\section*{Conflict of interest} On behalf of all authors, the corresponding author states that there is no conflict of interest.

	\newpage
 \bibliographystyle{apalike}
	\bibliography{bibliography}% common bib file

\end{document}